\documentclass[12pt]{article}
\usepackage{epsf,latexsym}
\usepackage[all]{xy}
\usepackage{amsfonts,amssymb} 
\usepackage{rotating}
\epsfverbosetrue
\newcommand{\de}{\hbox{\rm{d}}}

\newcommand{\bb}{\begin{eqnarray}}
\newcommand{\ee}{\end{eqnarray}}
\newcommand{\eee}{\nonumber\end{eqnarray}}
\newcommand{\qq}{\quad}

\begin{document}

\font\twelve=cmbx10 at 13pt
\font\eightrm=cmr8

\thispagestyle{empty}

\begin{center}
${}$
\vspace{3cm}

{\Large\textbf{Lensing in the Einstein-Straus solution. }} \\

\vspace{2cm}

{\large Thomas Sch\"ucker\footnote{also at Universit\'e de Provence, Marseille,
France, thomas.schucker@gmail.com } (CPT\footnote{Centre de Physique
Th\'eorique\\\indent${}$\qq\qq CNRS--Luminy, Case
907\\\indent${}$\qq\qq 13288 Marseille Cedex 9,
France\\\indent${}$\qq
Unit\'e Mixte de Recherche (UMR 6207) du CNRS et des Universit\'es
Aix--Marseille 1 et 2\\
\indent${}$\qq et Sud Toulon--Var, Laboratoire affili\'e \`a la
FRUMAM (FR 2291)})}

\vspace{2cm}

\vskip 2cm

{\large\textbf{Abstract}}
\end{center}
The analytical treatment of lensing in the Einstein-Straus solution with positive cosmological constant by Kantowski et al. is compared to the numerical treatment by the present author. The agreement is found to be excellent.

\vspace{2cm}

\vskip 1truecm

\noindent CPT-P031-2010\\
\vspace{1cm}

${}$

\section{Introduction}

Many applications of general relativity rely on  two solutions of Einstein's equation: (i) the outer Schwarzschild or -- in presence of a cosmological constant -- Kottler solution for tests in our solar system, (ii) the Friedmann solution used at cosmological scales. 
The Einstein-Straus solution \cite{es} merges both solutions. Such a joint solution is necessary for the understanding of weak and strong lensing because both are absent in Friedmann spaces for their symmetry. Also a naive superposition of Kottler's and Friedmann's solutions is incompatible with the non-linear nature of Einstein's equations. Ishak et al. \cite{ir} have used the Einstein-Straus solution to analyse the dependence of strong lensing on the cosmological constant. They present strong lensing in five clusters or galaxies including SDSS J1004+4112. More on this dependence can be found in the recent survey \cite{ir10}.  In reference \cite{t} you find a detailed numerical analysis with numbers concerning SDSS J1004+4112. Last year Kantowski et al. \cite{kant} have published an analytical formula for the bending angle of light in the Einstein-Straus solution and ZouZou et al. \cite{zz} just accomplished the computation of the time delay in the same situation. The aim of the present paper is a comparison between the numerical results of \cite{t} and the analytical result of Kantowski at al. \cite{kant}.

\begin{center}
\begin{tabular}{c}
\xy 
(0,0)*{}="L";
(-70,0)*{}="E";
(72,-6)*{}="S";
{\ar "E"; (78,0)*{}}; 
(0,0)*\xycircle(29,29){+\dir{-}};
(0,0)*\xycircle(25,25){};
"L"*{\bullet};
"E"*{\bullet};
"S"*{\bullet};
"E"; (-26,13) **\dir{-};
(-26,13); (18,26) **\dir{.};
(-26,13.1); (18,26.1) **\dir{.};
(-26,12.9); (18,25.9) **\dir{.};
(-34,17); "L" **\dir{.};
(-34,17.1); (0,0.1) **\dir{.};
(-34,16.9); (0,-0.1) **\dir{.};
{\ar "S"; (47.25,2.5)*{}};
(47.25,2.5); (22.5,11) **\dir{-};
(22.5,11); (-2.25,19.5) **\dir{.};
(22.5,11.1); (-2.25,19.6) **\dir{.};
(22.5,10.9); (-2.25,19.4) **\dir{.};
(0,0); (33.75,16.5) **\dir{.};
(0,0.1); (33.75,16.6) **\dir{.};
(0,-0.); (33.75,16.4) **\dir{.};
(-26,13) ; (22.5,11) **\crv{(-5,16)};
"L"; (1.5,13.5) **\dir{.};
(0.1,0); (1.6,13.5) **\dir{.};
(-0.1,0); (1.4,13.5) **\dir{.};
"E"; (-28,-15.8) **\dir{-};
(49.5,-10); (27,-14) **\dir{-};
{\ar "S"; (49.5,-10)*{}};
(-16,-17.3) ; (18,-15.8) **\crv{(3,-18)};
(-36,13.5)*{\gamma' _F};
(-32,16) ; (-32,11.1) **\crv{(-34,13)};
"L"; "S" **\dir{.};
(49,-2)*{-\varphi _S};
(44,0) ; (43.7,-3.7)**\crv{(45,-2)};
(37,12)*{\gamma' _{FS}};
(31,15) ; (31,8) **\crv{(34,11.5)};
(7,7.5)*{_{\pi /2 -\tilde\varphi _1'}};
(6,3) ; (1,5.7) **\crv{(4.5,6)};
(17,3)*{\varphi' _{\rm Sch\ddot u\, S}};
(9,4) ; (10,0) **\crv{(11,3)};
(-16.5,3)*{_{\pi -\varphi' _{\rm Sch\ddot u\, E}}};
(-8,4) ; (-9,0) **\crv{(-9.5,3)};
(11,18.5)*{\alpha'_{\rm tot} };
(5.5,17) ; (5.5,22.2) **\crv{(7,19.5)};
(-55.5,-2.5)*{\alpha };
(-59,0) ; (-59.9,-4) **\crv{(-58,-2)};
(-51,3)*{\alpha' };
(-55,0) ; (-55.8,4) **\crv{(-54,2)};
(-2,10.5)*{{r_P'}};
(74,-9)*{S};
(-73,-4)*{E};
(-4,-2.5)*{L};
(82,0)*{x};
(0,- 35)*{};
\endxy
\end{tabular}
\linebreak\nopagebreak
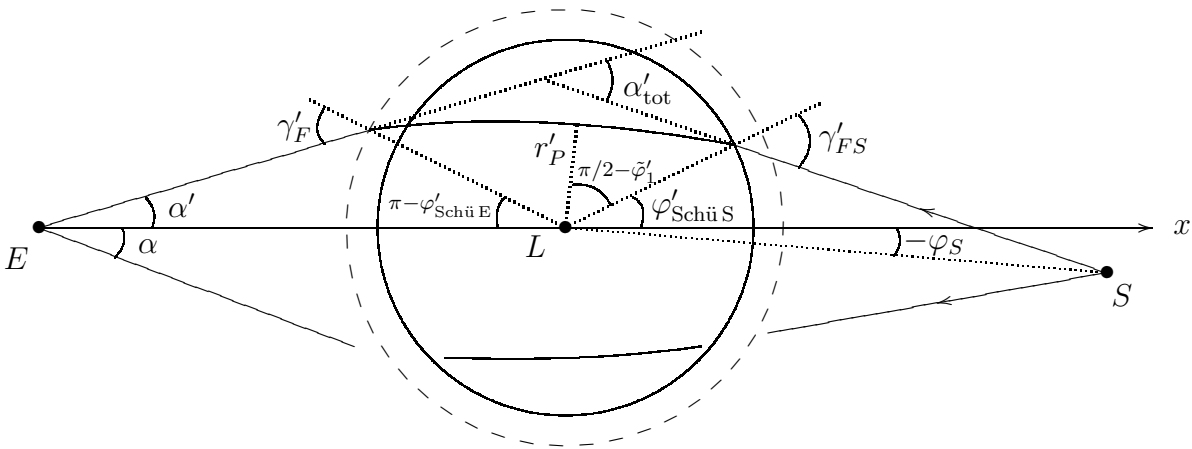
{Figure 1: The two light rays emitted from the source $S$ are refracted by the expanding Sch\"ucking sphere at four different radii and bent by the lens $L$ while inside the spheres.}
\end{center}

\section{The set up}

Let us summarise strong lensing in the Einstein-Straus solution using figure 1. Two light rays are emitted by the source $S$ and propagate  in Friedmann's metric along straight lines. They pass inside the expanding Sch\"ucking sphere at different times and radii and are refracted. While the rays are inside the Sch\"ucking sphere they are bent towards its center by the gravitational field of the concentrated, spherical lens $L$ of mass $M$ sitting at the center of the Sch\"ucking sphere. Upon exiting the Sch\"ucking sphere, again at different times and radii, the rays are refracted and then continue their trip on straight lines arriving at the Earth $E$ under angles $\alpha $ and $\alpha' $ with $\alpha' <\alpha $. The primed ray arrives  with a delay $\Delta t:=t '-t$. To avoid overcharging the figure, most details of the unprimed ray are suppressed.

Since inside the Sch\"ucking sphere we use the exterior Kottler solution, the following hierarchies of length scales must be satisfied at all times:
\bb s<r_{\rm cluster}<r_P<r_{\rm Sch\ddot u}(t)<D_{\rm cluster}/2\qq {\rm and}\qq r_{\rm Sch\ddot u}<
r_{\rm dS},\ee
where $s=2GM$ is
the Schwarzschild radius of the cluster, $r_{\rm cluster}$ the  radius of the cluster, $r_P$ the peri-cluster, $r_{\rm Sch\ddot u}(t)$ the Sch\"ucking radius as a function of cosmological  time $t$, $D_{\rm cluster}$ the typical distance between clusters and $r_{\rm dS} = (\Lambda /3)^{-1/2} $ is the de Sitter radius. Details are given in references \cite{t,kant}. 

\section{Comparison}
 
 The main result of Kantowski et al. \cite{kant} is an explicit perturbative formula giving the bending angle $\alpha _{\rm tot}$ as a function of the cosmological constant $\Lambda $, the lens mass $M$, the peri-cluster $r_P$ and the angle $\tilde\varphi _1$:
 \bb \alpha _{\rm tot}&=&\,\frac{s}{2r_P}\, \cos \tilde\varphi _1\left[ -4 \cos ^2 \tilde\varphi _1-12\cos \tilde\varphi _1\sin \tilde\varphi _1
 \sqrt{{\textstyle\frac{1}{3}} \Lambda r_P^2+\,\frac{s}{r_P}\, \sin ^3 \tilde\varphi _1}\right.\nonumber \\
 &&\qq\qq\qq\qq\qq\qq\qq\qq\qq\qq\qq\qq\qq\qq\qq\qq\qq\qq\qq\qq\qq\left. +\,{\frac{4}{3}} \,\Lambda r_P^2\left( 2-5 \sin ^2 \tilde\varphi _1\right) \right] \nonumber \\
 &&+\left( \frac{s}{2r_P}\right) ^2\left[ \,\frac{15}{4}\, (2\tilde\varphi _1-\pi)-12\log\left\{ \tan\,\frac{\tilde\varphi _1}{2}\, \right\} 
 \sin ^3 \tilde\varphi _1\right.
 \nonumber \\
 &&\qq\qq\qq\qq\qq\left.+\cos \tilde\varphi _1\left( 4+\,\frac{33}{2}\, \sin \tilde\varphi _1
 -4\sin^2 \tilde\varphi _1+19\sin^3 \tilde\varphi _1-64\sin^5 \tilde\varphi _1\right) \right]
 \nonumber \\
 &&\qq\qq\qq\qq\qq\qq\qq\qq\qq\qq\qq\qq\qq\qq\qq\qq\qq\qq\qq\qq\qq+\  \mathcal{O}\!\left( \,\frac{s}{r_P}\, +\Lambda r_P^2\right) ^{5/2}.\label{master}\ee
 This formula was derived under the assumption
 $s/r_P/\sin \tilde\varphi _1\ll 1$. Negative contributions to the bending angle are towards the lens. In principle the bending angle $\alpha _{\rm tot}$, the peri-cluster $r_P$ and the angle $\tilde\varphi _1$ are observable quantities, in practice they are not.

For concreteness,  let us consider the images $C$ and $D$ (primed quantities) of the lensed quasar SDSS J1004+4112 where the following quantities were observed \cite{in,fo}:
\bb \alpha =10''\,\pm\,10\% ,&
 z_L=0.68\ ,&M=5\cdot10^{13}M_\odot \pm20\% \ (r_{\rm cluster}=3\cdot10^{21} {\rm m}),
\\
\alpha ' =\ 5''\,\pm\,10\% ,& z_S=1.734,&\Delta t>5.7\ {\rm y\ (oct.\ '07)},
\ee
and let us use the spatially flat $\Lambda CDM$ model with $\Lambda = 1.36\cdot 10^{-52} \ {\rm m}^{-2}\ \pm 20\%$. In reference \cite{t} the mass of the cluster $M$ was computed numerically as a function of the cosmological constant and of the angles $\alpha $ and $\alpha '$ and using the measured redshift of the quasar $z_S$ and of the cluster $z_L$.  ZouZou et al. \cite {zz} have just published the time delay $\Delta t$ as a function of the same variables. We recollect these numbers in table 1. With respect to reference \cite{t}, table 1 has  higher precision and more intermediate variables: besides $\varphi _S,$ six others are exhibited, $ r_P,\,r'_P,\, \tilde\varphi _1,\,\tilde\varphi ' _1, \,\alpha _{\rm tot},\, \alpha' _{\rm tot}$. Note the correction found by ZouZou et al. \cite{zz}: the third mass value $1.7981\cdot 10^{13}M_\odot$ was wrongly reported as
 $1.7\cdot 10^{13}M_\odot$ in reference \cite{t}.
 
 The translation between the variables used by Kantowski et al. \cite{kant} and in reference \cite{t} are given by the following relations, which can be read from figure 1:
 \bb
 \alpha _{\rm tot}&=&\gamma _F+\gamma _{FS}+
 \varphi _{\rm Sch\ddot u\, S}-\varphi _{\rm Sch\ddot u\, E}-\pi ,\\[1mm]
 \alpha' _{\rm tot}&=&\gamma '_F+\gamma ' _{FS}+
 \varphi '_{\rm Sch\ddot u\, E}-\varphi ' _{\rm Sch\ddot u\, S}-\pi,\\[2mm]
  \tilde\varphi _1&=& \,\frac{\pi }{2}\,-\left(  \varphi _{\rm Sch\ddot u\, S}-\varphi _P\right) ,\\[2mm]
 \tilde\varphi' _1&=& \,\frac{\pi }{2}\,-\left( \varphi '_P- \varphi' _{\rm Sch\ddot u\, S}\right) ,
 \ee
where $ \varphi _{\rm Sch\ddot u\, S}-\varphi _P$ is obtained by integrating $\de\varphi /\de r$,
\bb
 \varphi _{\rm Sch\ddot u\, S}-\varphi _P&=&
 \,\frac{\pi }{2}\, -\arcsin\,\frac{r_P}{r_{\rm Sch\ddot u\, S}}\, 
 \nonumber\\[1mm]&&
 +{\textstyle\frac{1}{2}}  \,\frac{s}{r_{\rm Sch\ddot u\, S}}\,\sqrt{\,\frac{r_{\rm Sch\ddot u\, S}^2}{r_P^2}\,-1}+
{\textstyle\frac{1}{2}}\,\frac{s}{r_{\rm Sch\ddot u\, S}}\,\sqrt{\,\frac{r_{\rm Sch\ddot u\, S}-r_P}{r_{\rm Sch\ddot u\, S}+r_P}}
+\  \mathcal{O}\!\left( \,\frac{s}{r_P}\, \right),\\[2mm]
\varphi' _P-\varphi' _{\rm Sch\ddot u\, S}&=&
 \,\frac{\pi }{2}\, -\arcsin\,\frac{r'_P}{r'_{\rm Sch\ddot u\, S}}\, 
 \nonumber\\[1mm]&&
 +{\textstyle\frac{1}{2}}  \,\frac{s}{r'_{\rm Sch\ddot u\, S}}\,\sqrt{\,\frac{{r'}_{\rm Sch\ddot u\, S}^2}{{r'}_P^{2}}\,-1}+
{\textstyle\frac{1}{2}}\,\frac{s}{r'_{\rm Sch\ddot u\, S}}\,\sqrt{\,\frac{r'_{\rm Sch\ddot u\, S}-r'_P}{r'_{\rm Sch\ddot u\, S}+r'_P}}
+\,\mathcal{O}\!\left( \,\frac{s}{r'_P}\, \right).\ee
Finally the two columns  $\alpha  _{\rm tot\,K}$ and $\alpha'  _{\rm tot\,K}$ were computed using the explicit  formula (\ref{master}) by Kantowski et al.  For the indicated values we have: $s/r_P\sim 10^{-5}$, $\Lambda r_P^2\sim 10^{-9}$ and $s/r_P/\sin \tilde\varphi _1\sim 10^{-3}$ meeting the working assumptions of equation (\ref{master}). The agreement  between the numerical results, $\alpha  _{\rm tot}$ and $\alpha'  _{\rm tot}$, and the analytical results, $\alpha  _{\rm tot\,K}$ and $\alpha'  _{\rm tot\,K}$, is excellent.

\section{Conclusion}

The analytical formula (\ref{master}) by Kantowski et al. \cite{kant} for the bending angle is an important step towards  understanding how the cosmological constant modifies the bending of light. This understanding is precious for two reasons. 
\begin{itemize}\item
On the theoretical side, lensing in the Einstein-Straus solution is a concrete manifestation of the averaging problem, \cite{el,bu}. While the Einstein-Straus solution requires the same cosmological constant inside and outside the Sch\"ucking sphere, the central mass is `renormalised': $M=1.8\cdot10^{13}M_\odot$ calculated  from the  angles $\alpha $ and $\alpha '$ in the above example differs significantly from the value
$M=3.0\cdot10^{13}M_\odot$  obtained from the Kottler solution alone with a moving observer \cite{t}. Note also the non-monotonous dependence of $M$ on $\Lambda $.
\item
On the observational side, we are still looking for systems where these modifications are large enough with respect to the experimental uncertainties to be able to constrain the cosmological constant. 
\end{itemize}
Two other questions remain open.
\begin{itemize}\item
 In reality the lensed light rays pass through the galaxy or the cluster of galaxies, $r_P\not> r_{\rm cluster}$, see table 1 and $r_{\rm cluster}=3\cdot10^{21}\, {\rm m}$. Therefore we have to use an inner Kottler solution \cite{ik} inside the Sch\"ucking sphere.
 \item
  A generalisation of the above calculations to non-spherical lenses is still out of reach. 
\end{itemize}

\begin{sidewaystable}[h]
\begin{center}  
\begin{tabular}{|c|c|c||c|c||c||c|c||c|c||c|c|c|c|}
\hline
$\Lambda \pm  $&$\alpha \pm $&$\alpha' \pm $&$-\varphi _S$&$M$&$\Delta t $&$ r_P$&$ r'_P$&$\tilde\varphi _1$&$\tilde\varphi '_1$&$\alpha  _{\rm tot}$& $\alpha  _{\rm tot\,K}$& $\alpha'  _{\rm tot}$&$\alpha'  _{\rm tot\,K}$ \\ 
$ 20\% $&$10\%$&$  10\%$&$ ['']$&$ [10^{13}M_\odot]$&$[\rm{years}]$&$ [10^{21}\,{\rm m}]$&$  [10^{21}\,{\rm m}]$&$['']$&$['']$&$['']$& $ ['']$& $['']$&$['']$ \\ 
\hline\hline
$-$&$\pm 0$&$\pm 0$&
10.57&  1.8011&\ 9.53&
2.10205&1.05101&
5405.6&2694.6&
10.5589&10.5592&
21.1352&21.1355
                   \\ \hline
 $\pm0$&$\pm 0$&$\pm 0$&
\ 9.97&1.8200&\ 9.72&
2.25121&1.12559&
4862.9&2423.8&
9.96439&9.96462&
19.9420&19.9424
                   \\ \hline
 $+$&$\pm 0$&$\pm 0$&
 \ 9.03&{ 1.7981 }&\ 9.76&
 2.45579&1.22788&
 3682.1&1834.2&
 9.02778&9.02797&
 18.0625&18.0630
                   \\ \hline \hline
 $-$&$+$&$+$&
 11.63&2.1794&11.53&
 2.31225&1.15611&
 5580.1&2781.1&
 11.6141&11.6144&
 23.2485&23.2490
                   \\ \hline
$\pm 0$&$+$&$+$&
10.97&2.2022&11.76&
2.47633&1.23815&
5019.8&2501.5&
10.9606&10.9608&
21.9366&21.9372
                   \\ \hline
$+$&$+$&$+$&
\ 9.94&2.1757&11.81&
2.70137&1.35067&
3800.9&1892.9&
9.93038&9.93061&
19.8689&19.8695
                   \\ \hline \hline
 $-$&$+$&$-$&
 13.74&1.7831&12.41&
 2.31226&0.94591&
 5967.4&2431.2&
 9.50059&9.50081&
 23.2492&23.2496
                   \\ \hline
$\pm 0$&$+$&$-$&
12.97&1.8018&12.68&
2.47633&1.01303&
5368.3&2186.8&
8.96629&8.96648&
21.9371&21.9376
                   \\ \hline
$+$&$+$&$-$&
11.74&1.7801&12.77&
2.70138&1.10509&
4064.9&1654.5&
8.12411&8.12427&
19.8692&19.8698
                   \\ \hline\hline
$-$&$-$&$+$&
\ 7.40&1.7831&\ 6.60&
1.89184&1.15612&
4880.2&2976.2&
11.6168&11.6171&
19.0200&19.0203
                   \\ \hline
 $\pm 0$&$-$&$+$&
 \ 6.98&1.8018&\ 6.73&
 2.02608&1.23815&
 4390.2&2677.2&
 10.9626&10.9629&
 17.9470&17.9474
                   \\ \hline                   
$+$&$-$&$+$&
\ 6.32&1.7801&\ 6.74&
2.21021&1.35067&
3324.1&2026.2& 
9.93147&9.93170&
16.2558&16.2562
                   \\ \hline\hline
$-$&$-$&$-$&
\ 9.51&1.4588&\ 7.72&
1.89185&0.94591&
5219.1&2602.2&
9.50313&9.50334&
19.0207&19.0210
                   \\ \hline
$\pm 0$&$-$&$-$&
\ 8.98&1.4741&\ 7.87&
2.02609&1.01303&
4695.1&2340.7&
8.96826&8.96845&
17.9476&17.9479
                   \\ \hline
$+$&$-$&$-$&
\ 8.13&1.4564&\ 7.91&
2.21022&1.1051&
3555.0&1771.4&
8.12514&8.12530&
16.2561&16.2565
                   \\ \hline

\end{tabular}
\end{center}
\caption{  The  polar angle $\varphi _S$ between Earth and source  and the central mass $M$ are computed numerically as functions of the cosmological constant and of the measured angles $\alpha $ and $\alpha '$. `$\pm 0$' stands for the central value, '$+$' for the upper and `$-$' for the lower experimental limit. Other intermediate variables are reported, $r_P,\,r'_P,\, \tilde\varphi _1,\,\tilde\varphi ' _1, \,\alpha _{\rm tot},\, \alpha' _{\rm tot}$. The last two variables are also computed using the explicit formula (\ref{master}) by Kantowski et al. \cite{kant} yielding the values $\alpha  _{\rm tot\,K}$ and $\alpha'  _{\rm tot\,K}$. For completeness we reproduce the values for the time delay $\Delta t$ found by ZouZou et al. \cite{zz}.} 
\end{sidewaystable}


\begin{thebibliography}{10}

\bibitem{es}
A. Einstein and E. G. Straus,
``The influence of the expansion of space on the gravitation fields surrounding the individual star,''
Rev. Mod. Phys. {\bf 17} (1945) 120, {\bf 18} (1946) 148,
\\
E. Sch\"ucking,
``Das Schwarzschildsche Linienelement und die Expansion des Weltalls,''
Z.\ Phys.\  {\bf 137} (1954) 595.
\bibitem{ir}
  M.~Ishak, W.~Rindler, J.~Dossett, J.~Moldenhauer and C.~Allison,
  ``A New Independent Limit on the Cosmological Constant/Dark Energy from the
  Relativistic Bending of Light by Galaxies and Clusters of Galaxies,'' Mon. Not. R. Astron. Soc. {\bf 388} (2008) 1279
  [arXiv:0710.4726 [astro-ph]].
  \bibitem{ir10}
  M.~Ishak and W.~Rindler, 
  ``The Relevance of the Cosmological Constant for Lensing,'' 
 	arXiv:1006.0014 [astro-ph.CO],  Gen.\ Rel.\ Grav.\ in press.
 \bibitem{t}
  T.~Sch\"ucker,
  ``Strong lensing in the Einstein-Straus solution,''
  Gen.\ Rel.\ Grav.\  {\bf 41} (2009) 1595
  [arXiv:0807.0380 [astro-ph]].
\bibitem{kant}
  R.~Kantowski, B.~Chen and X.~Dai,
  ``Gravitational Lensing Corrections in Flat $\Lambda$CDM Cosmology,''
  arXiv:0909.3308 [astro-ph.CO].
  \bibitem{zz}
  K.~E.~Boudjemaa, M.~Guenouche and S.~R.~ZouZou,
  ``Time delay in the Einstein-Straus solution,''
  arXiv:1006.0080 [astro-ph.CO].
  \bibitem{in}
  N.~Inada {\it et al.}  [SDSS Collaboration],
  ``A Gravitationally Lensed Quasar with Quadruple Images Separated by 14.62
  Arcseconds,''
  Nature {\bf 426} (2003) 810
  [arXiv:astro-ph/0312427],\\
  M.~Oguri {\it et al.}  [SDSS Collaboration],
  ``Observations and Theoretical Implications of the Large Separation Lensed
  Quasar SDSS J1004+4112,''
  Astrophys.\ J.\  {\bf 605} (2004) 78
  [arXiv:astro-ph/0312429],\\
  N.~Ota {\it et al.},
  ``Chandra Observations of SDSS~J1004+4112: Constraints on the Lensing Cluster
and Anomalous X-Ray Flux Ratios of the Quadruply Imaged Quasar,''
  Astrophys.\ J.\  {\bf 647} (2006) 215
  [arXiv:astro-ph/0601700].
  \bibitem{fo}
  J.~Fohlmeister, C.~S.~Kochanek, E.~E.~Falco, C.~W.~Morgan and J.~Wambsganss,
  ``The Rewards of Patience: An 822 Day Time Delay in the Gravitational Lens
  SDSS J1004+4112,''
  arXiv:0710.1634 [astro-ph].
 \bibitem{el}
 G.F.R. Ellis, `` Relativistic cosmology Ð its nature, aims and problems,'' in General Relativity
and Gravitation (D. Reidel Publishing Co., Dordrecht), ed. B. Bertotti, F. de Felice
and A. Pascolini, pp. 215Ð288 (1984).
 \bibitem{bu}
T.~Buchert,
  ``Dark Energy from Structure - A Status Report,''
  Gen.\ Rel.\ Grav.\  {\bf 40} (2008) 467
  [arXiv:0707.2153 [gr-qc]].
\bibitem{ik}
  T.~Sch\"ucker,
  ``Lensing in an interior Kottler solution,''
  arXiv:0903.2940 [astro-ph.CO], Gen.\ Rel.\ Grav. in press.

 
\end{thebibliography}
\end{document}